\documentclass[twocolumn,showpacs,amsmath,amssymb,superscriptaddress]{revtex4}
\usepackage{dcolumn}
\usepackage{color}
\usepackage[dvips]{graphicx}
\usepackage{amsmath}
\usepackage{bm}

\begin{document}

\title{Shape coexistence in Lead isotopes in the interacting boson model
with Gogny energy density functional}

\author{K.~Nomura}
\email{nomura@ikp.uni-koeln.de}
\affiliation{Institut f\"ur Kernphysik, Universit\"at zu K\"oln, D-50937
K\"oln, Germany}

\author{R.~Rodr\'iguez-Guzm\'an}
\affiliation{Department of Physics and Astronomy, Rice University, Houston, Texas 77005,
USA} 
\affiliation{Department of Chemistry, Rice University, Houston, Texas 77005, USA}

\author{L.~M.~Robledo}
\affiliation{Departamento de F\'\i sica Te\'orica, Universidad
Aut\'onoma de Madrid, E-28049 Madrid, Spain}

\author{N.~Shimizu}
\affiliation{Center for Nuclear Study, University of Tokyo, Hongo,
Bunkyo-ku, 113-0033 Tokyo, Japan} 

\date{\today}

\begin{abstract}

We investigate the emergence and evolution of shape coexistence in
 the neutron-deficient Lead isotopes within the interacting boson model
 (IBM) plus configuration mixing with microscopic input based on the Gogny energy density
 functional (EDF). 
The microscopic potential energy surface obtained from the constrained 
 self-consistent Hartree-Fock-Bogoliubov method employing the Gogny-D1M
 EDF is mapped onto the coherent-state expectation value of the
 configuration-mixing IBM Hamiltonian. 
In this way, the parameters of the IBM Hamiltonian are fixed for each 
 of the three relevant configurations (spherical, prolate and oblate) 
 associated to the mean field minima. 
Subsequent diagonalization of the Hamiltonian provides the excitation 
energy of the low-lying states and transition strengths among them.
The model predictions for the  $0^{+}$ level energies and
 evolving shape coexistence in the considered Lead chain are consistent both
 with experiment and with the indications of the Gogny-EDF energy surfaces.

\end{abstract}



\maketitle

\section{Introduction}

The atomic nucleus is a physical system that exhibits a rich variety of intrinsic
geometrical shapes: spherical, prolate and oblate. 
The coexistence and evolution of the different intrinsic shapes has been a major theme
of interest in nuclear structure physics. It has been investigated
extensively from both theoretical and experimental sides in the past 
decades \cite{wood92,andre00,julin01rev,heyde11,duppen84,heyde85,heyde87}. 

In some specific regions of the Nuclide Chart, the energies of the three 
intrinsic geometrical shapes bunch together leading to the spectacular
coexistence of three $0^{+}$ states (including the ground state) in even-even nuclei.
Neutron-deficient Lead isotopes present a nice example of the shape 
coexistence phenomena \cite{andre00}: In the $^{186,188}$Pb nuclei, the presence
of three low-lying $0^{+}$ states and other additional
experimental data strongly suggest the 
coexistence of spherical, prolate and oblate shapes. 
In the context of the nuclear shell model \cite{federman77,heyde85,heyde87,heyde92}, the
emergence of low-lying excited $0^{+}$ states is traced back to the proton
particle-hole excitation across the $Z=82$ closed shell. 
The residual interaction between protons and neutrons is enhanced
due to this cross-shell excitation, resulting in the lowering of the excited $0^{+}$ states. 
In the vicinity of the $N=104$ mid-shell, the effect is strengthened 
and has a stronger impact on excitation energies. 
For the $^{186}$Pb nucleus, the three lowest $0^{+}$ states are within a
range of 700 keV and the two intruder
$0^{+}$ levels have the lowest excitation energy among the members of the Pb chain. 
The first excited $0^{+}$ state is interpreted as a proton
two-quasiparticle $(\pi h_{9/2})^{2}$ intruder configuration, while the 
second excited $0^{+}$ state could be interpreted as a  proton four-quasiparticle 
$(\pi h_{9/2})^{4}$ intruder configuration. 
These $0^{+}$ states correspond to oblate and  prolate equilibrium shapes. 

More quantitative results using large-scale shell model calculations can 
only be obtained in lighter nuclei. However, for heavy nuclei including the Lead 
isotopes the dimension of the shell model  configuration space becomes exceedingly large
and a truncation strategy preserving  the essential
ingredients of the low-energy spectrum is required. 
The Interacting Boson Model (IBM) \cite{IBM} has been successfully used for describing the
low-lying states of medium-heavy and heavy nuclei, and presents a severe
truncation of the full shell-model space \cite{Arima77,OAIT,OAI}. In this case, the building blocks are
$s$ and $d$ bosons, which reflect the collective
$J^{\pi}=0^{+}$ and $2^{+}$ pairs of valence nucleons, respectively \cite{Arima77,OAIT,OAI}. 

Within the IBM, the description of intruder $0^{+}$ states is based on the model by Duval and
Barrett \cite{Duval1981,Duval1982}. They proposed to mix the  normal ($0p$-$0h$) 
configuration, comprised of $N$ bosons, with intruder configurations comprised of $N+2n$
($n\geq 1,2,\ldots$) bosons, which takes into account the $2n$-particle-$2n$-hole excitation 
across the closed shell. In the case of Pb isotopes with three low-lying 
$0^{+}$ levels, the model consists of 
three different Hamiltonians corresponding to $0p$-$0h$, $2p$-$2h$
and $4p$-$4h$ configurations. 
The idea of configuration mixing in the IBM framework has been applied to spectroscopic 
analyses
\cite{Barfield83,heyde92,Fossion03,Hellemans08,Garcia11},
algebraic features \cite{DeCoster96,Lehmann97}, and geometry and phases \cite{frank04,frank06,morales08}
associated with the shape coexistence observed in the Lead and Mercury region. 
In these studies, the parameters for the configuration-mixing IBM
Hamiltonian have been extracted from a fit to the experimental spectra and
transition rates. 

The different configurations of the
shell model are related in the mean field language to the minima of the 
corresponding mean-field 
deformation energy surface. 
The self-consistent mean-field method using microscopic energy density
functionals (EDFs) currently provides an accurate and universal
description of nuclear ground-state properties and low-energy collective
excitations, including mass, density distributions, surface deformation,
giant resonance, etc. 
The most popular EDFs can be of zero-range Skyrme \cite{Ben03rev},
finite-range Gogny \cite{Gogny} 
as well as several parameterizations of the relativistic mean-field  (RMF)
Lagrangian \cite{vre05rev,Nik11rev}. The qualities and 
instabilities of the  self-consistent 
description of shape coexistence, based on a series of Skyrme 
interactions, were examined in \cite{Reinhard99}. 
On the other hand, the so called NLSC RMF parametrization has been 
tailored to describe the pronounced shape coexistence  
in Pb, Hg and Pt isotopes \cite{Nik02sc}.
The Nilsson-Strutinsky method has also been used to study the neutron-deficient
Pb and Hg isotopes \cite{Naza93}. 

At the mean-field level, however, important symmetries of the system
are spontaneously broken. Therefore, to describe the 
spectroscopic properties of a given
nucleus, one needs a systematic treatment of the dynamical effects 
associated with the restoration of the broken symmetries and fluctuations in the
collective coordinates. 
It is then necessary to project
the mean-field solutions onto states with good symmetry quantum numbers and mix the different configurations. 
Configuration mixing calculations,  in the spirit of the generator coordinate method
(GCM) have been performed for both Lead and Mercury nuclei, based on Skyrme \cite{Duguet03,Ben04Pb} and Gogny 
\cite{Chasman01,Rayner04Pb,egido04} EDFs. 

A sound approximation to the full GCM configuration mixing
calculation is represented by the solution of a five-dimensional 
collective Hamiltonian. 
Both vibrational and rotational mass parameters are obtained, from mean-field 
calculations, as
functions of the quadrupole collective variables. The collective potential is
then taken as the total energy surface resulting from the mean-field approximation from 
which, the zero-point energies associated with the rotational and vibrational
motions are subtracted \cite{Delaroche2010,Li10}. 
This method can  be also used  for the description of shape coexistence phenomena 
based on an arbitrary EDFs, e.g., using the Gogny-D1S functional for Hg
isotopes \cite{Delaroche94}.

More recently a comprehensive way of deriving the parameters of the 
IBM Hamiltonian has been introduced \cite{Nom08}. 
By mapping the potential energy surface, obtained within 
the constrained self-consistent mean-field 
method with a given EDF, onto the expectation value of the corresponding 
IBM Hamiltonian, the 
energy spectra and electromagnetic transition rates have been  computed. 
This method has been successfully applied to various shape phenomena, 
including vibrational and $\gamma$-unstable \cite{Nom10} as well as 
rotational deformed \cite{Nom11rot} nuclei, prolate-oblate shape
transitions \cite{Nom11pt} and to the study of the fingerprints of triaxiality \cite{Nom12}.

In this paper we extend the method of Ref.\cite{Nom08} to take into account 
configuration mixing within the IBM. We will show, how the parameters of the
configuration mixing IBM Hamiltonian can be determined without a fit
to the experiment by using the microscopic input provided by mean field
energy surfaces in an appropriate way. Using this method, we are able to 
describe the emergence and  evolution of shape coexistence. 
Our method  is applied to the neutron-deficient Pb isotopes since 
the existence of three minima in some of them represents a quite stringent 
test of the model. Moreover, they
are well studied both experimentally and theoretically offering us the possibility to 
benchmark our method with other proposals.
Concerning the mean-field calculation, we use the Gogny-D1M
\cite{D1M} functional that was originally fitted to binding
energies and radii. It has also shown good spectroscopic properties as already
exemplified in previous studies \cite{rayner10odd-1,rayner10odd-2,rayner10odd-3}
where it has been shown  that D1M keeps essentially the same predictive power 
as the standard Gogny-D1S EDF \cite{D1S}.

The paper is organized as follows: 
In Sec.~\ref{sec:theory},  a brief review of the 
configuration mixing within the  IBM and the geometrical interpretation is given.
The mapping of the microscopic PES to the IBM one with
configuration mixing is described and the way to extract the IBM parameters 
is discussed  in Sec.~\ref{sec:pes}.
In Sec.~\ref{sec:pb}, the results of the diagonalization of the IBM Hamiltonian 
including energy level systematics, the detailed level scheme
and the $B$(E2) transition strength values for
specific nuclei and the evolution of the 
spectroscopic quadrupole moment in the considered Pb chain are presented. 
Finally, Section \ref{sec:summary} 
is devoted to the conclusions and work perspectives.

\section{Description of the model}
\label{sec:theory}

\begin{figure*}[ctb!]
\begin{center}
\includegraphics[width=15cm]{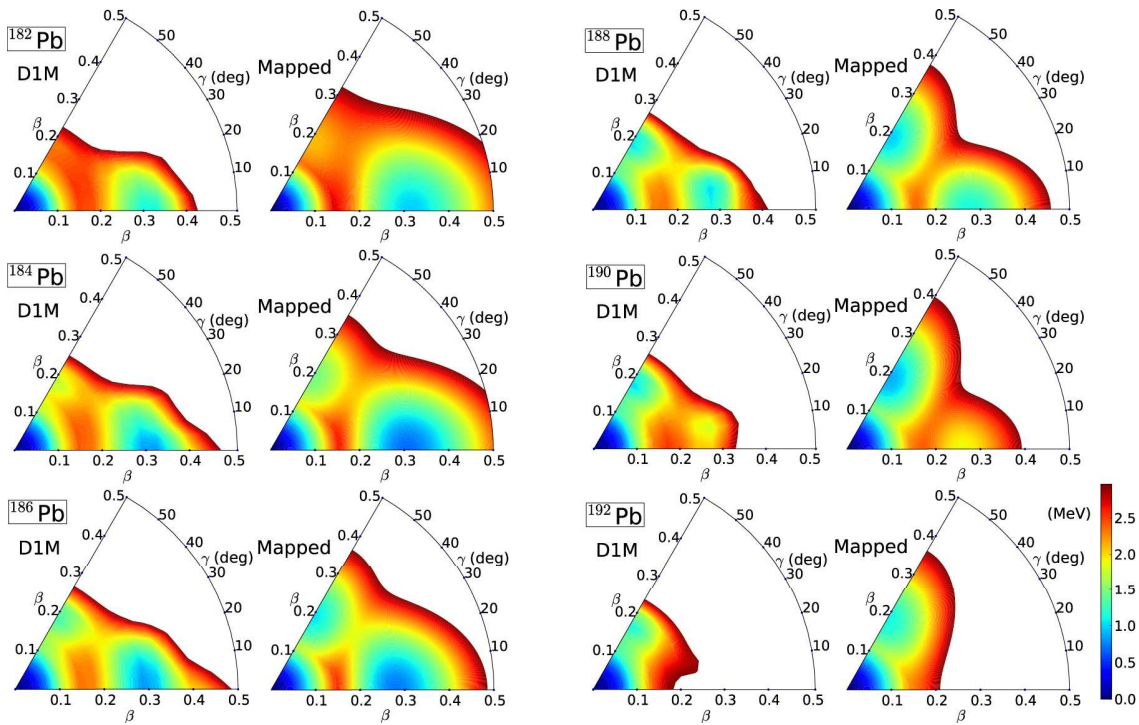}
\caption{(Color online) Contour plots of the PESs as a function of the
deformation parameters $\beta$ and $\gamma$ are given for the $^{182-192}$Pb isotopes.
The plots denoted by D1M correspond to the microscopic mean field calculation
with the Gogny-D1M EDF. The ones denoted by Mapped correspond to the 
mapped PESs used in fitting the IBM parameters. The color scale ranges
from zero (the mean field ground state) to 3 MeV.}
\label{fig:pes}
\end{center}
\end{figure*}

We start with the self-consistent constrained
Hartree-Fock-Bogoliubov (HFB) 
approximation based on the Gogny-D1M EDFs. As constraints we use the mass quadrupole moments 
associated with the
quadrupole deformation variables $\beta$ and $\gamma$ of the geometrical
collective model \cite{BM}. 
For a given set of  collective coordinate 
variables $q=(\beta, \gamma)$, HFB calculations are performed to obtain 
the potential energy surface (PES) given by the HFB
total energy denoted as $E^{\rm HFB}(\beta, \gamma)$ (for details the 
reader is referred to \cite{rayner10pt}).  
Note that, in some studies dealing with the five-dimensional 
collective Hamiltonian obtained from  EDF calculations \cite{Delaroche2010,Li10},
the PES is regarded as the total energy obtained after subtraction of the
rotational and vibrational zero-point energies to the HFB energy. In our model
the PES is simply the HFB energy  and no zero-point energy correction
are considered. A typical example of such PESs is
shown in Fig.~\ref{fig:pes} where the Gogny-D1M PESs in the ($\beta, \gamma$)
plane are given for the nuclei $^{182-192}$Pb. The Gogny-D1M EDF calculation produces a 
remarkable triple minima in $^{184-190}$Pb nuclei, where each local
minimum is well isolated from each other. 
In the considered Lead nuclei, the spherical minimum is always present
while a prolate minimum develops from $^{182}$Pb to $^{186}$Pb,
so does the oblate one. The prolate minimum becomes less significant
from $^{188}$Pb to $^{190}$Pb and finally disappears in $^{192}$Pb.
Within our model the Gogny-D1M EDF PES for an individual nucleus is
mapped onto the corresponding IBM PES (to be discussed below), as 
shown in Fig.~\ref{fig:pes} in the panels to the right of the Gogny ones. 

Let us turn to the IBM description of shape coexistence. 
In the present study, we consider the proton-neutron version of the
IBM (usually referred as IBM-2) \cite{Arima77,OAI} since it
takes into account  proton excitations more
explicitly than the original version of the IBM (IBM-1), which does
not distinguish between proton and neutron degrees of freedom.  
The IBM-2 comprises the neutron (proton) $s_{\nu}$ ($s_{\pi}$) and
$d_{\nu}$ ($d_{\pi}$) bosons, reflecting the neutron (proton) 
$J^{\pi}=0^{+}$ and $2^{+}$ collective pairs of valence nucleons \cite{Arima77,OAIT,OAI}.  
The number of neutron (proton) bosons, denoted as $N_{\nu}$ ($N_{\pi}$),
equals the number of neutron (proton) pairs outside the inert core. 

To describe a system consisting of three different intrinsic shapes, 
the Hilbert space is expressed as a direct sum of the
orthogonal subspaces for the normal ($0p$-$0h$) and the two intruder 
($2p$-$2h$ and $4p$-$4h$) configurations \cite{Duval1981,Duval1982}. 
The Hamiltonian of the system is written as 
\begin{eqnarray}
\hat H&=&\hat P_{0}\hat H_{0}\hat P_{0}+\hat P_{2}(\hat
 H_{2}+\Delta_{2})\hat P_{2}\nonumber \\
&&+\hat P_{4}(\hat
 H_{4}+\Delta_{4})\hat P_{4}+\hat H_{\rm
 mix}^{02}+\hat H_{\rm mix}^{24}
\label{eq:ham-tot}
\end{eqnarray}
where the $\hat H_{i}$ ($i=0,2,4$) represent the Hamiltonians for the  $ip$-$ih$ 
configurations associated with the different intrinsic shapes, 
$\hat H_{\rm mix}^{02}$ ($\hat H_{\rm mix}^{24}$) are the 
interaction terms mixing the $0p$-$0h$ ($2p$-$2h$) and the $2p$-$2h$ ($4p$-$4h$)
subspaces. The  operators $\hat P_{i}$ are projectors onto the $ip$-$ih$ configuration
spaces and finally the $\Delta_{i}$ ($i=2,4$) parameters represent the energies
needed to excite protons across the
$Z=82$ shell, which will be detailed later.

We employ the Hamiltonian $\hat H_{i}$ written as 
\begin{eqnarray}
 \hat H_{i}=\epsilon_{i}\hat n_{d}+\kappa_{i}\hat
  Q_{\pi}^{\chi_{\pi,i}}\cdot \hat
  Q_{\nu}^{\chi_{\nu,i}}, 
\label{eq:ham}
\end{eqnarray}
where the first term $\hat n_{d}=\hat n_{d\pi}+\hat n_{d\nu}$ represents the
$d$-boson number operator while  the second one is the  quadrupole-quadrupole
interaction between proton and neutron bosons. 
The quadrupole operator is defined as 
$\hat Q_{\rho}^{\chi_{\rho,i}}=s_{\rho}^{\dagger}\tilde
d_{\rho}+d_{\rho}^{\dagger}s_{\rho}+\chi_{\rho,i}[d_{\rho}^{\dagger}\tilde
d_{\rho}]^{(2)}$ ($\rho=\pi,\nu$). In this case $\epsilon_{i}$, $\kappa_{i}$ 
and $\chi_{\rho,i}$ are parameters. 
The Hamiltonian in Eq.~(\ref{eq:ham}) is taken in its simplified form in
order to reduce the number of parameters 
that are not directly determined from the PES. It keeps, however,
the essential aspects of a more general IBM-2 Hamiltonian. 

The mixing interaction terms $\hat H_{\rm mix}^{i-2\,i}$ ($i=2,4$) are defined as 
\begin{eqnarray}
 \hat H_{\rm
  mix}^{i-2\,i}
=\omega_{1}^{i-2\,i}(s_{\pi}^{\dagger}s_{\pi}^{\dagger}+s_{\pi}s_{\pi})
+\omega_{2}^{i-2\,i}(d_{\pi}^{\dagger}\cdot d_{\pi}^{\dagger}+\tilde
d_{\pi}\cdot\tilde d_{\pi}), 
\end{eqnarray}
where $\omega_{1}^{i-2\,i}$ and $\omega_{2}^{i-2\,i}$ stand
for the mixing strengths. 

In a shell model picture, the proton $2p$-$2h$ excitation across the closed
shell $Z=82$ creates one particle and one hole pairs in the $Z=82-126$
and the $Z=50-82$ major shells, respectively. 
Since the IBM normally does not distinguish
between particle and hole states, the $2n$-particle-$2n$-hole 
configuration comprises $2n$ additional proton bosons, and hence the
model contains
$N_{\nu}$ neutron bosons and $N_{\pi}+2n$ proton bosons.  
For the considered $^{182-192}$Pb nuclei, the 
doubly magic systems $^{164}$Pb and $^{208}$Pb
are assumed to be the inert cores. 
As a consequence, the proton boson numbers are $N_{\pi}=$0, 2 and 4 for regular, $2p$-$2h$
and $4p$-$4h$ configurations, respectively, while $N_{\nu}$ varies
between 8 and 11. 

A given IBM Hamiltonian can be related to the geometrical model by the
coherent-state framework \cite{GK}. 
The coherent state $|\Phi\rangle$ represents the intrinsic wave function of the boson
system, and is written, up to a normalization factor, as
\begin{eqnarray}
 |\Phi\rangle
=\prod_{\rho=\pi, \nu}
\Big(s^{\dagger}_{\rho}+\sum_{\mu=-2}^{2}a_{\rho\mu}d_{\rho\mu}^{\dagger}\Big)^{N_{\rho}}
|0\rangle
\label{eq:coherent}
\end{eqnarray}
where the coefficients $a_{\rho\mu}$ are given by 
$a_{\rho 0}=\beta_{\rho}\cos{\gamma_{\rho}}$, 
$a_{\rho\pm 1}=0$ and 
$a_{\rho\pm 2}=\frac{1}{\sqrt{2}}\beta_{\rho}\sin{\gamma_{\rho}}$. Here 
the parameters $\beta_{\rho}$ and $\gamma_{\rho}$ represent the axially-symmetric
and the triaxial deformations for neutrons ($\rho={\nu}$) and protons ($\rho={\pi}$), respectively. 
For simplicity we assume $\beta_{\nu}=\beta_{\pi}=\beta_{B}$ and
$\gamma_{\nu}=\gamma_{\pi}=\gamma_{B}$. 
The $\beta$ parameter for the IBM is proportional to the one in the geometrical model.
The proportionality coefficient is significantly larger than one  due to the difference in the
size of the model spaces \cite{GK}. On the other hand, the $\gamma$ variable can be
the same for the IBM and the geometrical model.   
The PES for the IBM system of interest is given analytically as an energy expectation value 
of the coherent state \cite{GK}. 

The geometrical interpretation of the configuration mixing IBM was 
provided by Frank {\it et al.} \cite{frank04}. 
The coherent state in Eq.~(\ref{eq:coherent}) for a single configuration
should be extended to be a direct sum of the coherent state for each configuration. 
The PES for the configuration mixing IBM is obtained as the lowest eigenvalue of the
following $3\times 3$ matrix \cite{frank04}
\begin{eqnarray}
 E(\beta,\gamma)=\left(
\begin{array}{ccc}
E_{0}(\beta,\gamma) & \Omega_{02}(\beta) & 0 \\
\Omega_{02}(\beta) & E_{2}(\beta,\gamma)+\Delta_{2} & \Omega_{24}(\beta) \\
0 & \Omega_{24}(\beta) & E_{4}(\beta,\gamma)+\Delta_{4} \\
\end{array}
\right), \nonumber \\
\label{eq:pes}
\end{eqnarray}
where the $E_{i}(\beta,\gamma)$ ($i=0,2,4$) in the diagonal part stands 
for the expectation value of the Hamiltonian $\hat H_{i}$ 
\begin{eqnarray}
&& E_{i}(\beta,\gamma)
=
\frac{\epsilon_{i}(N_{\nu}+N_{\pi,i})\beta_{B,i}^{2}}{1+\beta_{B,i}^{2}}
+\kappa_{i}N_{\nu}N_{\pi,i}\frac{\beta_{B,i}^{2}}{(1+\beta_{B,i}^{2})^{2}}\nonumber \\
&\times&\Big[
4-2\sqrt{\frac{2}{7}}(\chi_{\nu,i}+\chi_{\pi,i})\beta_{B,i}\cos{3\gamma}+\frac{2}{7}\chi_{\nu,i}\chi_{\pi,i}\beta_{B,i}^{2}
\Big]. 
\label{eq:pes-diag}
\end{eqnarray}
Here $\beta_{B}^{i}=C_{\beta,i}\beta$, with $C_{\beta,i}$ being the
proportionality coefficient of the $\beta$ variable defined for the
different mean-field minima associated with each configuration
$ip$-$ih$, and $N_{\pi,i}$ denotes the proton boson number in the $ip$-$ih$ configuration. 
The non-diagonal entries $\Omega_{i-2\,i}(\beta)$ ($i=2,4$) represent the expectation values of 
the mixing interactions $\hat H_{\rm mix}^{i-2\,i}$, given as 
\begin{eqnarray}
 \Omega_{i-2\,i}(\beta)
&=&
\frac{\sqrt{N_{\pi,i}(N_{\pi,i}-1)}}{1+\beta_{B,i}^{2}}(\omega_{1}^{i-2\,i}+\omega_{2}^{i-2\,i}\beta_{B,i}^{2})\nonumber \\
&\times&
\Big(
\frac{1+\beta_{B,i-2}\beta_{B,i}}{\sqrt{(1+\beta_{B,i-2}^{2})(1+\beta_{B,i}^{2})}}
\Big)^{N_{\nu}+N_{\pi,i-2}}. 
\label{eq:pes-nondiag}
\end{eqnarray}

Each of the microscopic PES, presented in Fig.~\ref{fig:pes},
is mapped onto the corresponding IBM PES, i.e., the lowest eigenvalue of
the matrix in Eq.~(\ref{eq:pes}). 
Since the three local minima are well separated from each other, a set of 
parameters for each configuration are determined independently from each others. 
First, the $0p$-$0h$ configuration is assigned to the mean-field minimum with the smallest deformation. 
Then the $2p$-$2h$ configuration is assigned to the minimum
with second larger quadrupole deformation. 
Likewise the $4p$-$4h$ configuration is associated with the minimum with the
third larger quadrupole deformation. 
For each configuration, the parameters $\epsilon_{i}$,
$\kappa_{i}$, $\chi_{\nu,i}$, $\chi_{\pi,i}$ and $C_{\beta,i}$ in
$E_{i}(\beta,\gamma)$ of Eq.~(\ref{eq:pes-diag}) are determined, using the method of 
Ref.~\cite{Nom10}, so that the topologies, i.e., curvatures in both $\beta$ and 
$\gamma$ directions, around the corresponding minima are reproduced. 
For $^{186}$Pb, for instance, the Hamiltonians for $0p$-$0h$,
$2p$-$2h$ and $4p$-$4h$ configurations are assigned to spherical ($\beta=0$), 
oblate ($\beta\approx -0.2$) and prolate ($\beta\approx +0.3$) minima, respectively. 
Since the number of proton bosons $N_{\pi}$ is zero for all the considered Pb nuclei, 
the second term in Eq.~(\ref{eq:ham}) vanishes,
and the parameters $\kappa_{0}$, $\chi_{\nu,0}$ and 
$\chi_{\pi,0}$ can be set to zero. 
Therefore, in the present study, the $0p$-$0h$ configuration always represents a pure U(5) limit
of the IBM \cite{IBM}. 

The $\Delta_{i}$ parameters in Eq.~(\ref{eq:pes}) are constants 
depending on the nucleus and they are fixed so that the energy difference 
between the mean-field spherical and intruder configurations 
is reproduced.  These energy differences between  mean-field minima are  
denoted as 
$\delta E_{i}=E^{\rm HFB}(\beta^{i}_{\rm min},\gamma^{i}_{\rm min})-E^{\rm
HFB}(\beta^{0}_{\rm min},\gamma^{0}_{\rm min})$ 
with ($\beta^{i}_{\rm min},\gamma^{i}_{\rm min}$) being the coordinates
that give the minimum for each of the $ip$-$ih$ configuration in the HFB PES. 
These quantities should be in reasonable agreement with the 
observed $0^{+}$ excitation energies.

However, the values of the $\Delta_{i}$ derived
from the estimation above should not be used
in the spectroscopic calculations with the Hamiltonian of 
Eq.~(\ref{eq:ham-tot}), i.e., different values of  $\Delta_{i}$ should be used
in Eqs.~(\ref{eq:ham-tot}) and (\ref{eq:pes}). 
From the original definition, the $\Delta_{i}$ ($i=2\,{\rm or}\,4$) represents the offset energy 
added to the eigenenergies of the $ip$-$ih$ Hamiltonian so 
that its ground-state $0^{+}$ energy exceeds that of the normal configuration by 
an amount that is roughly equal to the observed excited $0^{+}$ energy 
and hence to $\delta E_{i}$. 
More explicitly (cf. Appendix C of \cite{Duval1982}), 
\begin{eqnarray}
 E_{i}(0^{+})+\Delta_{i}=E_{0}(0^{+})+\delta E_{i}, 
\label{eq:delta}
\end{eqnarray}
where $E_{i}(0^{+})$ represents the lowest (ground-state) $0^{+}$ eigenvalue of 
the $ip$-$ih$ Hamiltonian in Eq.~(\ref{eq:ham}). 
Note that the amount of energy gained by the mixing between normal and
intruder configurations is much
smaller than the typical range of $\Delta_{i}$ values and is considered negligible in this rough estimate. 
In the considered Pb isotopes, since there is no
deformation-driving term in the Hamiltonian $\hat H_{0}$,
the $E_{0}(0^{+})$ energy is always equals to zero for the $0p$-$0h$ configuration.  
The lowest $0^{+}$ eigen-energy 
comprises the energy gained through the deformation at the mean-field
level (equivalent to the depth of the minimum in the PES) and the extra correlation
energy arising from quantum effects beyond the mean field.  
The $\Delta_{i}$ values determined solely by  looking at the PES, 
do not take into account this quantum correlation energy, and hence is too
small to describe correct spectroscopic tendencies  consistent
with the indications of the microscopic PESs. 

Let us consider, for example, the nucleus $^{186}$Pb. 
The $\Delta_{2}$ value derived from the PES, to be used in 
Eq.~(\ref{eq:pes}), is 4.014 MeV. 
Nevertheless, with this value, the intruder
$0^{+}$ state becomes ground state after the mixing. 
This is apparently not consistent with empirical facts as well as with
the indication of the microscopic PES. 
Since the $2p$-$2h$ configuration gives $E_{2}(0^{+})=-3.676$ MeV, to reproduce $\delta E_{2}=1.208$ MeV 
the $\Delta_{2}$ value to be used in Eq.~(\ref{eq:ham-tot}) should amount to 
$\Delta_{2}=1.208-(-3.676)=4.884$ MeV. 
The difference between the two $\Delta_{2}$ values ($=0.870$ MeV), identified as the
quantum correlation energy that the $2p$-$2h$ configuration gains through the
diagonalization, seems so sizable as to change the conclusion. 
Therefore, for the spectroscopic calculations with the Hamiltonian of Eq. 
(\ref{eq:ham-tot}), we propose to use the formula in
Eq.~(\ref{eq:delta}) to take into account the necessary quantum
correlation effects. 
Also the $\Delta_{i}$ in Eq.~(\ref{eq:ham-tot}) can be related to the
ones in Eq.~(\ref{eq:pes}) by replacing
$E_{i}(0^{+})$ in Eq.~(\ref{eq:delta}) by the deformation energy
$E_{i}(\beta^{i}_{\rm min},\gamma^{i}_{\rm min})$, and vice versa. 
The uncertainty in the parameters relevant to the configuration mixing 
has also been pointed out in Ref.\cite{frank04}, where the PES of 
the configuration mixing IBM-1 Hamiltonian for Lead nuclei was analyzed. 
Although the parameters of the Hamiltonian give a good description of
the spectroscopy, only two (spherical and prolate) minima remain 
after configuration mixing in the $^{186}$Pb nucleus \cite{frank04}. 
This result seems to support our finding  that the $\Delta_{i}$ values to
be used in spectroscopic calculations may not at the same time give 
the IBM mapped PES similar in topology to the mean-field PES. 

To perform a fully consistent mapping of $\Delta_{i}$ in the present
framework, the addition of some interaction term between like neutron bosons, 
such as of $\kappa_{\nu}\hat Q_{\nu}\cdot\hat Q_{\nu}$ type, to the $0p$-$0h$ 
Hamiltonian may solve the problem. The reason is that such term drives deformation
and provides the energy which could compensate for the quantum correlation energy
the intruder configuration gains. 
In fact, if one tries to put $\kappa_{\nu}\hat Q_{\nu}\cdot\hat Q_{\nu}$
with the realistic interaction strength $\kappa_{\nu}=-0.013$ MeV in the
mapped Hamiltonian $\hat H_{0}$ in Eq.~(\ref{eq:ham}) for $^{186}$Pb, 
the $0p$-$0h$ configuration gives $E_{0}(0^{+})=-0.870$ MeV, which is 
exactly the same as the correlation energy gained in the $2p$-$2h$ configuration. 
Nevertheless, since the microscopic Gogny-D1M PES suggests purely
spherical minimum for the normal configuration, it is practically not
possible to determine the strength parameter for such additional
interaction term. Another possible solution which could work out is to map the
angular-momentum projected PES onto the corresponding IBM PES. 
This could represent an interesting work for the future which is out of the scope 
of the present paper. 

The non-diagonal matrix elements, $\Omega_{i-2\,i}(\beta)$ in
Eq.~(\ref{eq:pes-nondiag}), concern the barrier between the mean-field
minima but are only minor as compared to the diagonal parts in Eq.~(\ref{eq:pes-diag}). 
Therefore, the parameter $\omega^{i-2\,i}$ can be introduced only
perturbatively and is determined so that the barrier
height for two different minima in the microscopic PES is reproduced. 
For the sake of simplicity, we assume
$\omega_{1}^{i-2\,i}=\omega_{2}^{i-2\,i}\equiv\omega^{i-2\,i}$.  

\section{Mapped IBM potential energy surfaces and derived parameters}
\label{sec:pes}

\begin{figure*}[ctb!]
\begin{center}
\includegraphics[width=16.0cm]{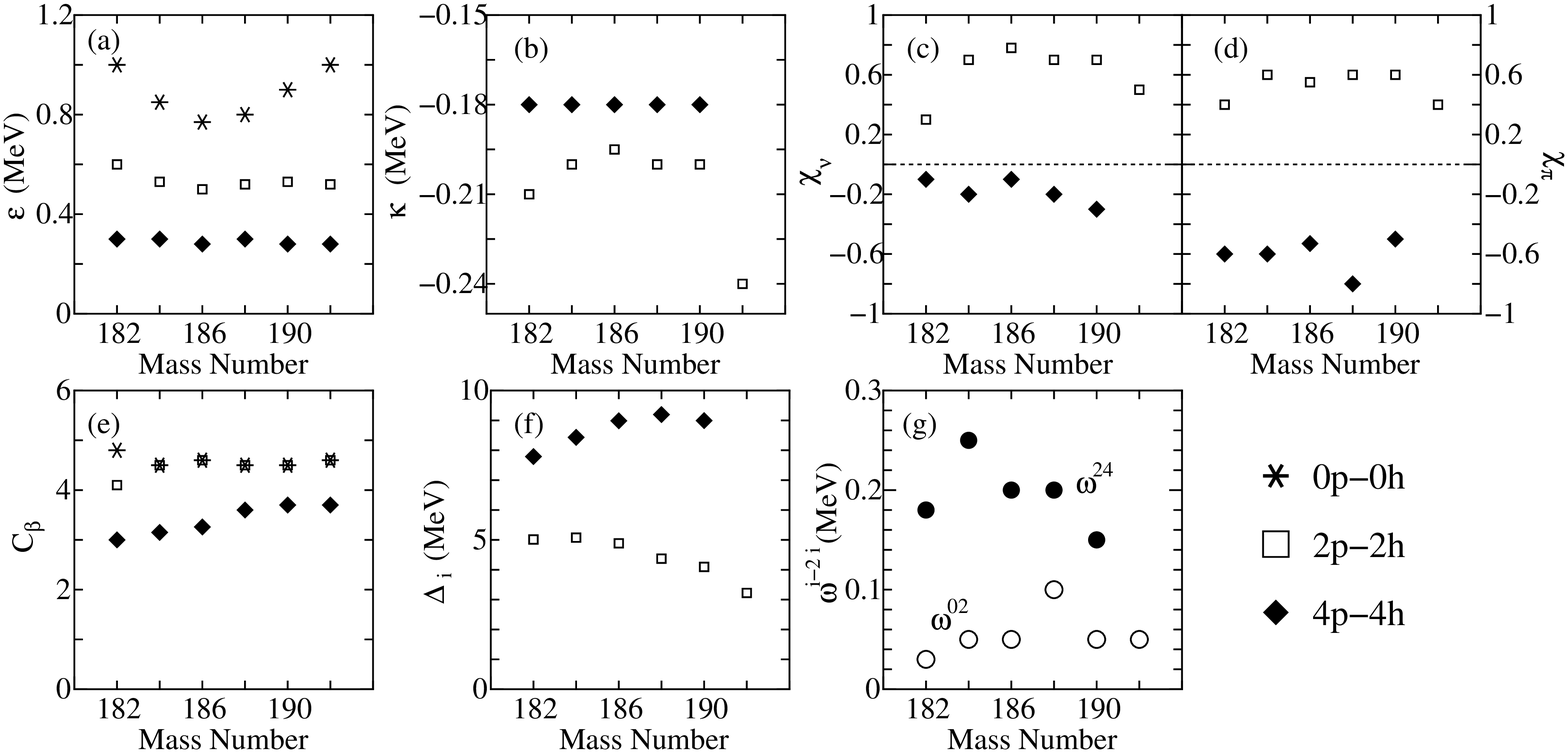}
\caption{Derived IBM parameters (a) $\epsilon_{i}$, (b) $\kappa_{i}$,
 (c) $\chi_{\nu,i}$, (d) $\chi_{\pi,i}$, (e) $C_{\beta,i}$, (f)
 $\Delta_{i}$ and (g) $\omega^{i-2\,i}$ for the considered $^{182-192}$Pb nuclei as  functions
 of mass number A. Figure legends for (a) through (f) are shown in the right hand side of panel (g). }
\label{fig:para}
\end{center}
\end{figure*}

The mapped IBM and the microscopic Gogny-D1M HFB PESs are plotted in
Fig.~\ref{fig:pes} for the nuclei $^{182-192}$Pb.
In the case of $^{192}$Pb the HFB approximation suggests two minima, and
therefore only the $0p$-$0h$ and $2p$-$2h$ configurations are mixed in this
nucleus. The location, relative energy differences as well as the 
energy barriers between the coexisting minima in the microscopic 
PESs are reproduced rather well in
the mapped IBM PESs. Note that, due to the limited number of bosons,
the mapped PESs are generally flat along the oblate axis.
Although  very shallow triaxial minima at $\gamma\approx 10^{\circ}$
are displayed in the 
HFB PESs of $^{188,190}$Pb, in the mapped IBM PESs such minima 
are approximated by axial ones. As a result, some deviations 
of the barrier heights between the oblate and prolate minima
occur for these nuclei. In order to describe the 
detailed energy systematics of quasi-$\gamma$ band, a boson 
three-body term \cite{Nom12} is required which is, however, 
out of the scope of the present work.
 
The IBM parameters, derived for the considered isotopes  $^{182-192}$Pb, are
displayed in Fig.~\ref{fig:para}. Consistent with the evolution 
of the topology in the PESs shown  in Fig.~\ref{fig:pes}, no rapid
change with mass number is observed in these 
parameters. The comparison between the $\epsilon$ parameters, for a given nucleus, in  Fig.~\ref{fig:para}(a)
 reveals that  $\epsilon_{0}$  is the  largest, $\epsilon_{4}$  is the smallest
while the $\epsilon_{2}$ value is always in
between them. On the other hand, as a function of the mass number, 
$\epsilon_{0}$ looks parabolic with respect to the mid-shell
nucleus $^{186}$Pb, while $\epsilon_{2}$ and $\epsilon_{4}$ remain almost constant. 
Let us stress, that these boson 
number dependencies are consistent with the earlier phenomenological
(see Ref.~\cite{IBM} and references are therein) and microscopic
\cite{OAIT,Mizusaki97} IBM-2 studies on collective structural evolution. 
The parameter $\kappa_{2}$ is, in general, larger  than  $\kappa_{4}$ as  
the model
space of the latter contains a larger number of bosons. 
As  functions of the mass number the $\chi$ 
parameters, shown in Figs.~\ref{fig:para}(c) and
 \ref{fig:para}(d), also display a weak dependence.  Nevertheless, the sign 
of $\chi_{\nu}$ is always opposite to the one of 
$\chi_{\pi}$.  Their sum $\chi_{\nu}+\chi_{\pi}$ is positive (negative) for the 
oblate (prolate)  $2p$-$2h$ ($4p$-$4h$) shapes. 
The $C_{\beta}$ value (i.e., the scale factor for the $\beta$ variable) does
not change too much. This parameter is determined from the position of
the axial minimum and the curvature along the $\beta$ axis. 
Actually, the location of each axial minimum in the HFB PES in
Fig.~\ref{fig:pes} remains almost the same in the corresponding IBM PES. 

Figures~\ref{fig:para}(f) and \ref{fig:para}(g) show the energy offset
$\Delta_{i}$, defined in Eq.~(\ref{eq:delta}), and the mixing parameters for the 
Hamiltonian $\hat H^{i-2\,i}_{\rm mix}$, respectively. 
The magnitudes of both $\omega^{02}$ and $\omega^{24}$ are notably
larger than those used in some fitted calculations within the configuration-mixing
IBM-1 model \cite{Fossion03,Hellemans08}: $\omega^{02}\approx 10$ keV and
$\omega^{24}\approx 20-30$ keV in the latter studies, while we have obtained
$\omega^{02}\approx 50$ keV and $\omega^{24}\approx 200$ keV. 
The present $\omega^{24}$ value, which is particularly larger than the
one derived from phenomenology,  implies that our microscopic EDF
approximation suggests a complex topology of the mean-field PESs in the
studied Lead isotopes in $\gamma$ direction. 
Therefore, it may require a mixing between the two intruder
configuration spaces stronger than estimated from the pure fitting calculations. In 
particular, the mixing between the regular and  $2p$-$2h$ configurations seems to be
quite large in  the case of  $^{188}$Pb. 

The offset energy $\Delta_{i}$, depicted in Fig.~\ref{fig:para}(g), roughly
amounts to 4 and 8 MeV for the $2p$-$2h$ and $4p$-$4h$ configurations, respectively. These values are 
approximately twice as large as the ones obtained 
in the IBM-1 phenomenology \cite{Hellemans08}. 
One sees from Eq.~(\ref{eq:delta}), that a  larger $\Delta_{i}$ energy is needed 
when the $0^{+}$ eigenenergy of the intruder configuration is sufficiently
large in magnitude compared to the $0^{+}$ energy of the normal configuration.
The quadrupole-quadrupole interaction for the intruder configuration
appears to be stronger in the present mapped IBM system than it is in
the IBM-1 phenomenology. 
The intruder configuration gains a large amount of energy, giving rise to
remarkable differences between our $\Delta_{i}$ values and the 
phenomenological \cite{Fossion03,Hellemans08} results. 
In fact, the derived $\kappa_{2}$ and $\kappa_{4}$
values are larger in magnitude than those extracted from the fit. 
This may be due to the fact that the microscopic Gogny-D1M 
calculation (see, Fig.~\ref{fig:pes}) provides a pronounced minimum. 

\section{Spectroscopic results}
\label{sec:pb}

Having determined all the parameters required by the IBM Hamiltonian in 
Eq.~(\ref{eq:ham-tot}) for each individual nucleus, the energy spectra and
transition rates are calculated by diagonalizing the Hamiltonian within
the enlarged model space consisting  of the direct sum of 
the $0p$-$0h$, the $2p$-$2h$ and the $4p$-$4h$ subspaces. 
The results shown below are obtained without any fit to the
experimental data, but only from the  Gogny-D1M HFB approximation 
and the mapping procedure described above. 

We have performed a diagonalization of the mapped IBM-2 Hamiltonian in the so
 called boson $m$-scheme basis. 
 The eigenfunction for each excited state gives rise to various spectral
observables. In particular, the E2 transition rates and the
spectroscopic quadrupole moments are important quantities by which one
can gauge the emergence and the evolution 
of the coexistence and competition between different  shapes  
in the considered isotopes. 
For the E2 operator $\hat T^{({\rm E2})}$, we use the boson quadrupole
operator $\hat Q_{\rho,i}$, where the same parameter $\chi_{\rho,i}$ as
 the one used in diagonalization is used, based on the idea of Casten
 and Warner for the IBM-1 case \cite{War83}. 
Within the configuration mixing IBM framework this E2 operator can be written as \cite{Duval1981,Duval1982}
\begin{eqnarray}
 \hat T^{({\rm E2})}=\sum_{\rho,i}e_{\rho,i}\hat P_{i}\hat
  Q_{\rho,i}^{\chi_{\rho,i}}\hat P_{i}, 
\label{eq:E2}
\end{eqnarray}
where $e_{\rho,i}$ represents the proton and neutron boson effective charges for
each configuration. 
For simplicity, these charges are assumed to be the 
same (i.e., $e_{\nu,i}=e_{\pi,i}\equiv e_{i}$). 
For the effective charges, we have adopted  the values 
given in Ref.~\cite{Hellemans08} (i.e., $e_{0}=0.110$, $e_{2}=0.140$ and
$e_{4}=0.170$ eb). 
The effective charge should, in principle, be determined by taking into
account core polarization effects. Such an effect could be renormalized 
in the effective charges used here, while
a fully microscopic derivation of the boson effective charge still represents an
interesting open problem. With all this in mind, the reduced E2 transition 
$B({\rm E2};J\rightarrow J^{\prime})$ between states with spins 
$J$ and $J^{\prime}$, can be written as 
\begin{eqnarray}
 B({\rm E2};J\rightarrow J^{\prime})=\frac{1}{2J+1}|\langle
  J^{\prime}||\hat T^{({\rm E2})}||J\rangle |^{2}, 
\end{eqnarray}
where $|J\rangle$ and $|J^{\prime}\rangle$ represent the wave functions
of the initial and the final states with angular
momenta $J$ and $J^{\prime}$, respectively. 

The spectroscopic quadrupole moment $Q^{(s)}(J)$ for the state with spin $J$
is given by 
\begin{eqnarray}
 Q^{(s)}(J)=\sqrt{\frac{16\pi}{5}}
\left(
\begin{array}{ccc}
J & 2 & J \\
-J & 0 & J \\
\end{array}
\right)
\langle J||\hat T^{({\rm E2})}||J\rangle, 
\label{eq:qm}
\end{eqnarray}
where use is made of the well known Wigner's 3-j symbol \cite{varshalovich1988quantum}.

\subsection{Level-energy systematics}

\begin{figure*}[ctb!]
\begin{center}
\includegraphics[width=14.0cm]{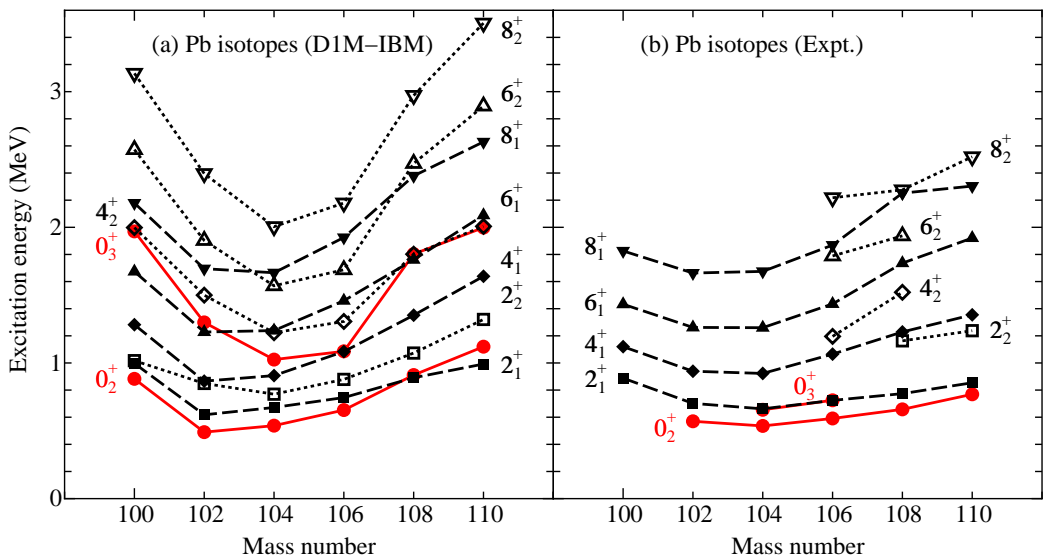}
\caption{(Color online) Level-energy systematics for $^{182-192}$Pb
 isotopes with mass number. Theoretical level energies resulting from the mapped IBM-2
 Hamiltonian (a) are compared with the experimental (b) energies. The
 experimental data are taken from the ENSDF database 
 \cite{data}. To guide the eye, each point has been connected. Solid,
 dashed and dotted lines stand for the lowest two excited $0^{+}$ states, yrast states with 
 $J\geq 2$ ($2^{+}_{1}$, $4^{+}_{1}$, $6^{+}_{1}$ and $8^{+}_{1}$)
 and non-yrast states with
 $J\geq 2$ ($2^{+}_{2}$, $4^{+}_{2}$, $6^{+}_{2}$ and
 $8^{+}_{2}$), respectively. Note that the experimental $2^{+}_{1}$ and $0^{+}_{3}$
 excitation energies for $^{186}$Pb ($^{188}$Pb) are 662 (724) keV and
 655 (725) keV, respectively. }
\label{fig:level}
\end{center}
\end{figure*}

Figure~\ref{fig:level} displays the theoretical (a) and the experimental
\cite{data} (b) low-lying spectra as functions of the mass number. 
In the nuclei $^{184-188}$Pb, the relative location of the $0^{+}_{2}$
and $2^{+}_{1}$ experimental levels is nicely reproduced. Our 
calculations  reproduce the correct location in energy for
 these first excited
$0^{+}$ states, with the $0^{+}_{2}$ level coming down 
as we approach the midshell nucleus $^{186}$Pb and becoming the lowest-energy
one at $^{186}$Pb or $^{188}$Pb. Both prolate and oblate minima 
become lowest in energy for these nuclei (see, Fig.~\ref{fig:pes}) 
and therefore the residual quadrupole-quadrupole correlation between neutron bosons
and the intruder proton bosons become maximal, giving rise to 
these notably low-lying excited $0^{+}$ states.

The comparison between our results and the few available 
data for the excitation energy of the $0^{+}_{3}$
states reveals that our values overestimate the experimental
ones. This could be due to the fact that in  the considered
isotopes, the third lowest-energy minimum in the mean-field 
PESs appears  higher than expected from the
experimental point of view and also because of the level repulsion. 
 Note, that the parabolic behaviour of the 
 $0^{+}_{3}$ levels with respect to mid-shell is in good agreement
 with
the relative location of the three minima in the 
Gogny-PESs (see, Fig.~\ref{fig:pes}): the 
 three minima are closest  to each other around 
 $^{186}$Pb while 
the second and third minima become
less pronounced and only the spherical one remains 
as we 
approach the closed shells.

The present calculations also reproduce the parabolic tendency of 
states with angular momenta $J\geq 2$. The collectivity of
 the intruder configurations becomes stronger and, as a result, the 
intruder states with $J\geq 2$ become most compressed around the midshell.  
Nevertheless, the change in all the calculated energy levels, including
the excited $0^{+}$ ones, takes place faster as compared with the
experimental trend. Let us also stress that, similar to 
the situation observed for the $0^{+}_{3}$ levels, the calculated non-yrast 
$2^{+}_{2}$,
$4^{+}_{2}$, $6^{+}_{2}$ and $8^{+}_{2}$ spectra are 
more stretched than the experimental ones.

\subsection{Structure of eigenfunctions}

\begin{table}[cb!]
\caption{\label{tab:frac} Fraction of each configuration in the lowest
 three $0^{+}$ states of the considered $^{182-192}$Pb isotopes (in \%). }
\begin{center}
\begin{tabular}{c c c c c c c c}
\hline\hline
\textrm{$J^{\pi}$}&
\textrm{Configurations}&
\textrm{$^{182}$Pb}&
\textrm{$^{184}$Pb}&
\textrm{$^{186}$Pb}&
\textrm{$^{188}$Pb}&
\textrm{$^{190}$Pb}&
\textrm{$^{192}$Pb}\\
\hline
            & $0p$-$0h$ & 100.0 & 99.8 & 99.7 & 98.6 & 99.7 & 99.6 \\
$0^{+}_{1}$ & $2p$-$2h$ &   0.0 &  0.2 &  0.2 &  1.3 &  0.3 &  0.4 \\
            & $4p$-$4h$ &   0.0 &  0.0 &  0.1 &  0.1 &  0.0 &   -  \\
\hline
            & $0p$-$0h$ &   0.0 &  0.1 &  0.1 &  1.4 &  0.5 &  0.5 \\
$0^{+}_{2}$ & $2p$-$2h$ &   9.0 & 14.6 & 24.3 & 65.5 & 92.8 & 99.5 \\
            & $4p$-$4h$ &  91.0 & 85.3 & 75.6 & 33.1 &  6.7 &   -  \\
\hline
            & $0p$-$0h$ &  34.1 &  0.9 &  0.5 &  0.7 & 98.7 & 93.2 \\
$0^{+}_{3}$ & $2p$-$2h$ &  41.2 & 67.9 & 67.5 & 36.1 &  1.2 &  6.8 \\
            & $4p$-$4h$ &  24.7 & 31.2 & 32.0 & 63.2 &  0.1 &   -  \\
\hline\hline
\end{tabular}
\end{center}
\end{table}

To interpret the dominant component in the calculated excited states and
the structure of the  wave functions, we show in
 Table ~\ref{tab:frac}, the 
 overlap probabilities of the basis
states and the eigenfunctions corresponding to the
 three lowest-excited $0^{+}$ states for all the considered
Lead isotopes. In all the  isotopes the
$0^{+}_{1}$ state corresponds to the spherical ground state
with a nearly 100 \% dominance of 
the $0p$-$0h$ configuration. For the nuclei $^{182,184,186}$Pb, the first excited $0^{+}$ state is comprised
predominantly of the $4p$-$4h$ configuration, which corresponds to the prolate
minimum in Fig.~\ref{fig:pes}. 
The extent of mixing between the $2p$-$2h$ and the $4p$-$4h$
configurations for the first excited $0^{+}$ state becomes gradually
stronger from  $^{182,184}$Pb to  $^{186}$Pb, which
correlates well with the finding in Fig.~\ref{fig:pes} that the oblate minimum becomes more
significant from $^{182,184}$Pb to $^{186}$Pb.

Experimentally both $^{186,188}$Pb are regarded as the  most spectacular
examples of shape
coexistence in the Pb isotopic chain. In this case, one sees a stronger
 mixing between different configurations in the first
and the second excited $0^{+}$ states. The $0^{+}_{2}$ state in  $^{186}$Pb is
more or less clearly
of $4p$-$4h$ character while the two intruder configurations are mixed
for the $0^{+}_{2}$ state in $^{188}$Pb. The earlier IBM-1 fitting 
calculation \cite{Fossion03} suggested almost the same
predominance of the $0^{+}_{2}$ and the $0^{+}_{3}$ eigenfunctions
while the three configurations appear to be
more strongly mixed for $^{186}$Pb. On the other hand, the present 
results for $^{186}$Pb  seem to be consistent with 
the ones obtained within the symmetry-projected GCM approximation
based on both the 
Skyrme-SLy6 \cite{Duguet03} and Gogny-D1S \cite{Rayner04Pb} EDFs.
In such studies \cite{Duguet03,Rayner04Pb}, the collective wave function for the
$0^{+}_{2}$ ($0^{+}_{3}$) excited state is peaked  
on the prolate  (oblate) side. We also find, that our results
for the nucleus $^{188}$Pb agree qualitatively well with the ones 
of previous symmetry-projected GCM studies 
\cite{Rayner04Pb,Ben04Pb} where collective wave functions 
strongly peaked at the oblate and prolate sides have also been 
predicted. For the nuclei  $^{190,192}$Pb, there is almost no 
mixing between the different configurations for the three
$0^{+}$ states. In fact, the fraction of the $4p$-$4h$ configuration 
is too small for them. 

\subsection{Level scheme: $^{186,188}$Pb nuclei}

\begin{figure}[ctb!]
\begin{center}
\includegraphics[width=8.6cm]{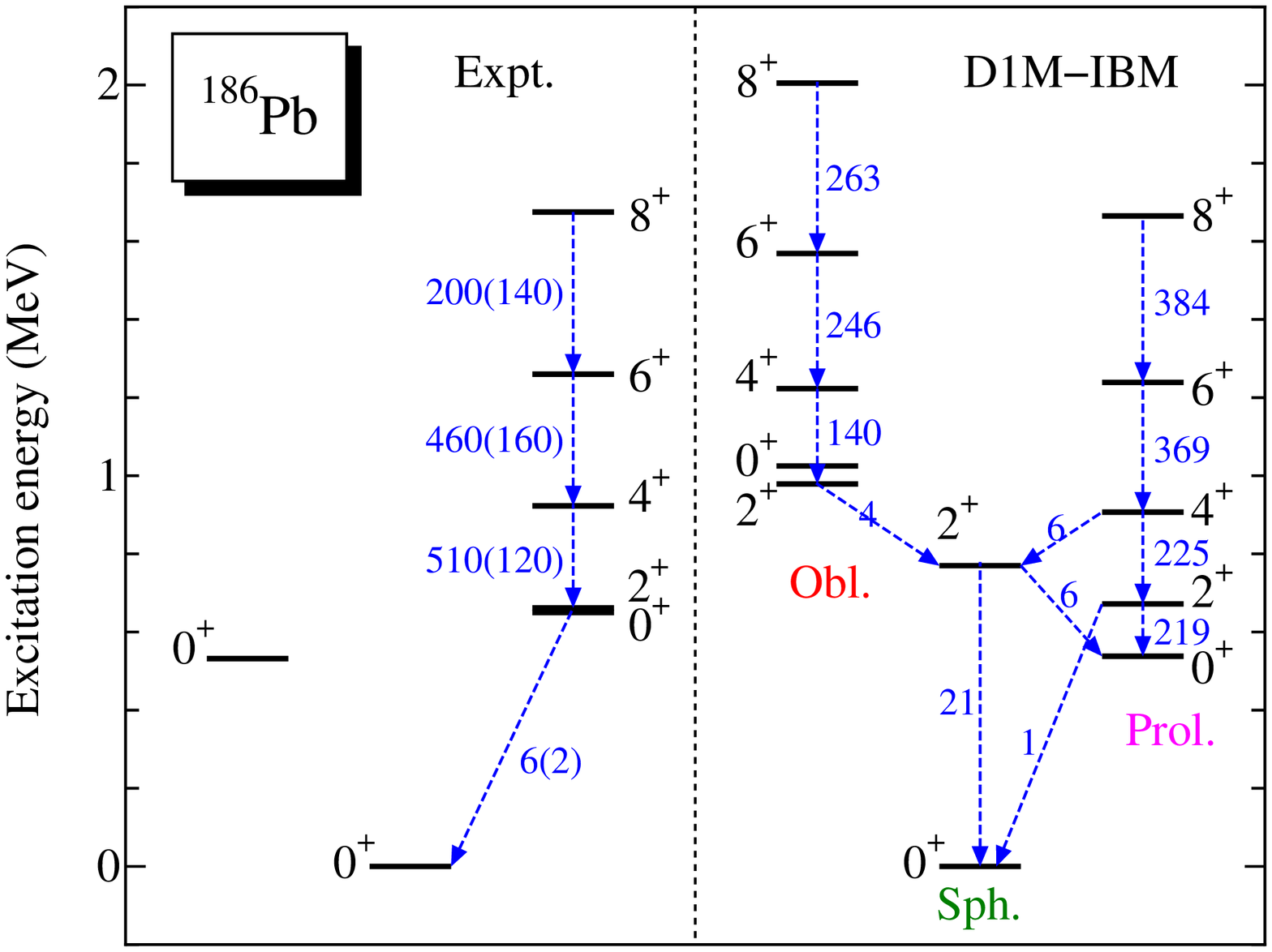}
\caption{(Color online) Experimental and calculated energy spectra and
 $B$(E2) transition rates (in Weisskopf units) for the $^{186}$Pb
 nucleus. Experimental energies and $B$(E2) values are taken from 
 \cite{data,Grahn06}. In the plot the experimental $0^{+}_{3}$ and $2^{+}_{1}$
 look nearly degenerated, but their excitation energies are 650 keV
 and 662 keV, respectively. The experimental $B({\rm E2})$ of 510(120) and
 6 (2) (W.u.) correspond to the $4^{+}_{1}\rightarrow 2^{+}_{1}$ and 
 $2^{+}_{1}\rightarrow 0^{+}_{1}$ transitions, respectively. }
\label{fig:pb186}
\end{center}
\end{figure}

\begin{figure}[ctb!]
\begin{center}
\includegraphics[width=8.6cm]{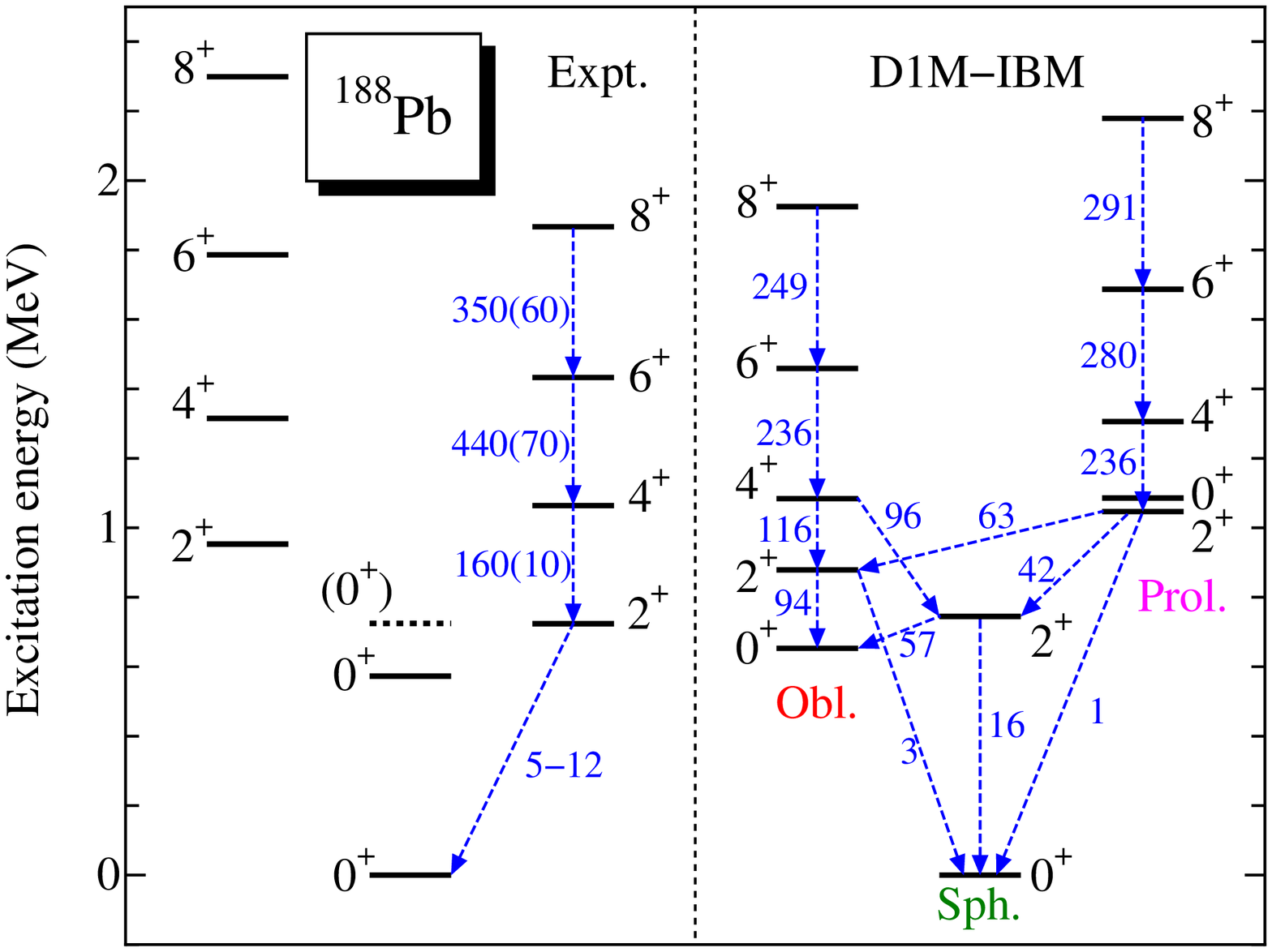}
\caption{(Color online) Same as Fig.~\ref{fig:pb186}, but for the $^{188}$Pb
 nucleus. Note that the theoretical $0^{+}_{3}$ and $2^{+}_{3}$ excitation
 energies are 1.086 MeV and 1.047 MeV, respectively. The following theoretical
 $B({\rm E2})$ in the right panel are listed here to help identify the 
 corresponding transition in the plot:
 $B({\rm E2};4^{+}_{2}\rightarrow 2^{+}_{3})=236$, 
$B({\rm E2};2^{+}_{3}\rightarrow 2^{+}_{2})=63$, 
$B({\rm E2};2^{+}_{3}\rightarrow 2^{+}_{1})=42$ and $B({\rm
 E2};2^{+}_{3}\rightarrow 0^{+}_{1})=1$ (in W.u.).}
\label{fig:pb188}
\end{center}
\end{figure}

In this section, we discuss in more detail the results obtained
for the isotopes $^{186}$Pb and
$^{188}$Pb which are the most  distinct cases of shape
coexistence in the considered chain. We compare in Figs.~\ref{fig:pb186} 
and \ref{fig:pb188} our theoretical 
and the experimental energy levels and transition rates for these nuclei. 
The assignment of the calculated excited state to each band is 
done according to the predominance of a given configuration in the
corresponding eigenstate and the E2 transition strength that exhibits a
clear collectivity. 

For $^{186}$Pb, in Fig.~\ref{fig:pb186}, the
calculated  first
excited $0^{+}_{2}$ state, predicted to be predominantly 
prolate, is quite close to the experimental value. From 
the experimental point of view, such state 
has been identified \cite{andre00} as the oblate band head. 
On the other hand, our result in Fig.~\ref{fig:pb186} is 
consistent with 
earlier predictions for the same nucleus within the 
symmetry-projected GCM approximation based on the functionals
Skyrme-SLy6 \cite{Duguet03} and Gogny-D1S \cite{Rayner04Pb}.
Actually, as seen from Table~\ref{tab:frac}, the $4p$-$4h$ (prolate in the
present IBM framework) component 
dominates 75.6 \% of the $0^{+}_{2}$ state.
A strong collective energy pattern is also predicted for this
prolate band, with the ratio $\Delta E_{ 4^{+}_{1}}/\Delta E_{ 2^{+}_{1}}=2.75$. 
 The $B$(E2) transitions among the members of this prolate band exhibit
 a collective behavior while the $2^{+}\rightarrow 0^{+}$ E2 transition is
very weak in the spherical band. Concerning the oblate 
band, the theoretical $0^{+}_{3}$ excitation
energy overestimates the experimental one. 
Note that, experimentally, this $0^{+}_{3}$ state is recognized as the prolate
bandhead \cite{andre00}. The experimental $2^{+}_{1}$ and $0^{+}_{3}$ levels look nearly
degenerated, and so does the present calculation except that the
$2^{+}_{3}$ level lies slightly below the $0^{+}_{3}$ level since the
mixing between the two intruder configurations may be too strong. 

One notices from Fig.~\ref{fig:pb188}, that our model
provides a similar level of quality in the description of the isotope
$^{188}$Pb. 
Although the calculated excitation energy for the $0^{+}_{3}$ state
 is a bit high, the calculated
$0^{+}_{2}$ state lies close to the experimental one. 
The present study also suggests, that the $0^{+}_{2}$ and the $0^{+}_{3}$
levels correspond to oblate and prolate configurations, respectively, which is 
consistent with symmetry-projected GCM calculations based on 
the Gogny-D1S EDF \cite{Rayner04Pb}. 
Nevertheless, the first and the second excited $0^{+}$ states are
experimentally \cite{data} interpreted as the
prolate and the oblate bandheads, respectively. 
Moreover, the present study suggests a 
pronounced collective pattern for both the prolate ($4p$-$4h$) and the
oblate ($2p$-$2h$) bands, and supports the experimental evidence for the
strong E2 transition pattern in the
band comprised of $2^{+}_{1}$, $4^{+}_{1}$, $6^{+}_{1}$,
and $8^{+}_{1}$ states. 
 In our calculations, the two intruder $0^{+}$ levels are rather
close in energy, compared to the case of  $^{186}$Pb.
In fact, among all the considered nuclei, the prolate-oblate 
energy difference obtained from the 
Gogny-D1M PESs in Fig.~\ref{fig:pes} reaches its lowest value
for  $^{188}$Pb. Due to the level repulsion, however, the 
excitation energy of the
$0^{+}_{3}$ state 
is larger than the energy difference between the spherical and prolate minima 
of the corresponding HFB PES in Fig.~\ref{fig:pes}. 

\subsection{Spectroscopic quadrupole moment}

\begin{figure}[ctb!]
\begin{center}
\includegraphics[width=7.0cm]{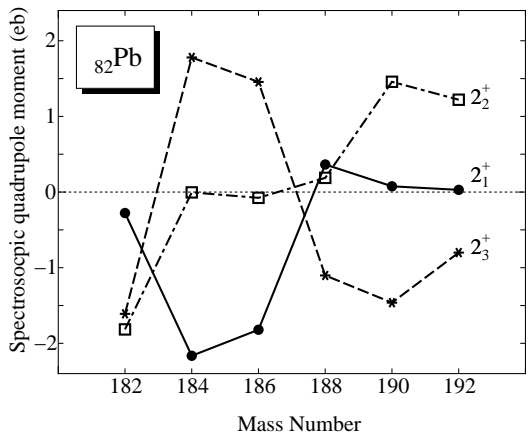}
\caption{Calculated spectroscopic quadrupole moments $Q^{(s)}$ for the lowest
 three excited $2^{+}$ states of the considered Pb nuclei as functions
 of mass number. Solid, dashed, and dot-dashed lines connect the
 calculated $Q^{(s)}$ values for $2^{+}_{1}$, $2^{+}_{2}$ and
 $2^{+}_{3}$ states, respectively. }
\label{fig:qm}
\end{center}
\end{figure}

The spectroscopic quadrupole moment $Q^{(s)}$, computed according 
to  Eq.~(\ref{eq:qm}), is shown in Fig.~\ref{fig:qm} as a function
of the mass number for the  three lowest  $2^{+}$ excited 
states of the considered
Lead isotopes. In the case of 
$^{182}$Pb, $Q^{(s)}(2^{+}_{1})\approx 0$ eb and
$Q^{(s)}(2^{+}_{2})\approx Q^{(s)}(2^{+}_{3})$ $\approx$ 
-2 eb,  reflecting the spherical and prolate character 
of the $2^{+}_{1}$ and $2^{+}_{2},2^{+}_{3}$ states, respectively. 
The microscopic and the mapped PESs for this nucleus
(see, Fig.~\ref{fig:pes}) suggest a
global spherical minimum  and a well developed  prolate deformation.
The two non-yrast $2^{+}$ states should originate from such a pronounced
prolate minimum. For $^{184,186}$Pb, both the HFB and 
the mapped PESs in Fig.~\ref{fig:pes} 
indicate the development of triple coexistence.
The trend of the considered quadrupole moment changes accordingly. 
The $Q^{(s)}(2^{+}_{1})$ ($Q^{(s)}(2^{+}_{3})$) value is nearly $-2$
($+1.5$) eb, suggesting that this state is 
prolate (oblate). From Fig.~\ref{fig:pb186}, one 
realizes that the prolate band consisting
of the   $0^{+}_{2}, 2^{+}_{1}, 4^{+}_{1}, 6^{+}_{1}\,{\rm and}\,8^{+}_{1}$
states comes down in energy. On the other hand, our 
calculations suggest that the third band in 
$^{186}$Pb, comprised of the  $2^{+}_{3}$, $0^{+}_{3}$, $4^{+}_{2}$, 
$6^{+}_{2}$, and $8^{+}_{2}$ states, originates from the $2p$-$2h$ oblate configuration.
Note, that the quadrupole moment for the $2^{+}_{3}$ state
is positive. The same arguments apply to the nucleus $^{184}$Pb. 

A change in the spectroscopic quadrupole moments 
from $^{186}$Pb to $^{188}$Pb
is also apparent from 
Fig.~\ref{fig:qm}. In $^{188}$Pb, $Q^{(s)}(2^{+}_{2})=0.19$ eb
while $Q^{(s)}(2^{+}_{1})=0.36$ eb. On the other hand, the
$Q^{(s)}(2^{+}_{3})$ value becomes negative ($=-1.10$ eb). 
As can be observed from the level scheme displayed 
in Fig.~\ref{fig:pb188}, the
$2^{+}_{1}$ state consists exclusively of the regular (spherical)
configuration. The band consisting of the  
$0_{2}^{+}, 2^{+}_{2}, 4^{+}_{1}, 6^{+}_{1}$ and $8^{+}_{1}$ states
emerges
with predominant $2p$-$2h$ oblate character while the one
comprised of the
$0^{+}_{3}, 2^{+}_{3}, 4^{+}_{2}, 6^{+}_{2}\,{\rm and}\,8^{+}_{2}$ 
states emerges with $4p$-$4h$ prolate character.
It should be noted, however, that 
the spectroscopic quadrupole moment $Q^{(s)}(2^{+}_{2})$ for 
the $2^{+}_{2}$ state, assigned  
to the oblate band due to its 
stronger E2 transition to the 
$0^{+}_{2}$ state, is quite close to the 
$Q^{(s)}(2^{+}_{1})$ value. 
This is mainly due to the fact that the mixing 
between the different configurations is
too strong for these two $2^{+}$ states: for the $2^{+}_{1}$ ($2^{+}_{2}$) 
state, 58 (41), 32 (38) and 10 (21) \% of its
eigenfunction is composed of spherical $0p$-$0h$, oblate $2p$-$2h$ 
and prolate $4p$-$4h$ configurations, respectively.
 The value $Q^{(s)}(2^{+}_{3})=-1.10$ eb  reflects a more clear prolate
character, as the three configurations are less strongly mixed in this 
$2^{+}_{3}$ state: 2, 38 and 60 \% of the eigenfunction is composed of $0p$-$0h$, $2p$-$2h$ 
and $4p$-$4h$ configurations, respectively. 

For both $^{190,192}$Pb, we obtain that $Q^{(s)}(2^{+}_{1})$ is 
close to zero so that the $2^{+}_{1}$ state is supposed to 
to be of spherical character. 
Our result seems to
support the fact that the $2^{+}_{2}$ state is 
composed predominantly of the
$2p$-$2h$ oblate configuration.  
This result agrees well with 
the corresponding PESs, shown in Fig.~\ref{fig:pes}, for which
the oblate minimum lies 
much lower, compared to $^{182-188}$Pb, than the prolate one.
Note also that $Q^{(s)}(2^{+}_{3})<0$
for $^{190,192}$Pb implying, that the $2^{+}_{3}$ state is prolate.

\section{Summary}
\label{sec:summary}

To summarize, the emergence and  evolution of the shape coexistence in the neutron-deficient
Lead isotopes have been investigated within the 
configuration mixing IBM model with parameters extracted solely from
a mapping of the mean-field PESs obtained with the  
Gogny-D1M EDF. The diagonalization of the IBM Hamiltonian provides
energy levels as well as 
 transition rates between the excited states. 
It is important to emphasize that,
although the IBM configuration mixing model contains many parameters, they
can be determined unambiguously by relating the IBM PES for each 
configuration to the corresponding mean-field deformation minimum in 
the microscopic PES. No additional adjustment to experimental data is required.
A potential difficulty and uncertainty of the fully consistent mapping 
concerning the offset energy $\Delta$ has been addressed and possible remedies for it have
been discussed. 

The considered Lead nuclei present the most spectacular example of the
coexistence of spherical, oblate and prolate equilibrium shapes.
The relative location of the three associated $0^{+}$ states were reproduced. 
In one of the most stringent tests, the $^{186}$Pb nucleus, the present
calculation suggested that the $0^{+}_{2}$ and the $0^{+}_{3}$ states
are predominantly of prolate ($4p$-$4h$) and oblate ($2p$-$2h$) nature, respectively. 
For the $^{188}$Pb nucleus, another typical example with more
available experimental data to compare with, the present work predicts the oblate bandhead as the 
first excited $0^{+}$ state and the prolate band as the second excited $0^{+}$ state. 
The calculated E2 transition pattern, albeit the quantitative deviation 
of the inter-band transitions from the experimental data, provides 
indications of strong collectivity for the relevant prolate and oblate 
shapes. 
The experimental level-energy systematics is well reproduced by our calculations. 
The study of the prolate-oblate dynamics has been complemented by looking at
the spectroscopic quadrupole moment. Its value for different configurations
and nuclei is consistent
with the implications of other quantities and the suggestions of the mean-field
 microscopic calculations. 

Using the proposed methodology, many new 
research directions concerning complex shape dynamics are opened up. 
A possible application would be to analyze neighboring isotopic chains, Mercury, Polonium and Platinum
isotopes. In particular, the study in the Platinum isotopes will help to
disentangle if the single-configuration is the appropriate picture to describe those 
isotopes (see, e.g., Ref.~\cite{Garcia11,Nom11pt} and references are therein). 
Other mass regions, including neutron-deficient krypton,
selenium and germanium isotopes, and neutron-rich krypton, strontium and
zirconium isotopes, which are also known as regions of 
shape coexistence \cite{heyde11} would be a potential target. 

The predictive power endowed to the model by the microscopic input makes 
possible the application of the present methodology  to the study of 
exotic nuclei like the ones that will be experimentally accessible in 
the near future.

\begin{acknowledgments}
Authors would like to thank J . Jolie and R. V. Jolos for valuable discussions,
 and T. Otsuka for his continuous interest in this work. 
Author K. N. acknowledges the support by the JSPS Postdoctoral
 Fellowships for Research Abroad. 
LMR acknowledges support of MINECO through grants Nos. FPA2009-08958 and
FIS2009-07277 as well as the Consolider-Ingenio 2010 program CPAN 
CSD2007-00042 and MULTIDARK  CSD2009-00064.
Author N. S. acknowledges the support by SPIRE field 5, MEXT, Japan.
\end{acknowledgments}

\bibliography{refs}

\end{document}